\documentclass[newstyle,twocolumn,proceedings]{rmaa}

\title{A Catalogue of IJK Photometry of PNe with DENIS}

\suppressfulladdresses
\author{S. Schmeja, S. Kimeswenger \affil{Institut f\"ur Astrophysik, Universit\"at Innsbruck, Technikerstr.\ 25, A-6020 Innsbruck, Austria\\
 stefan.j.schmeja@uibk.ac.at, stefan.kimeswenger@uibk.ac.at} }

\shortauthor{Schmeja \& Kimeswenger}
\shorttitle{IJK Photometry of PNe}

\listofauthors{S. Schmeja \& S. Kimeswenger}
\indexauthor{Kimeswenger, S.}
\indexauthor{Schmeja, S.}

\begin{document}
\maketitle

\noindent\textbf{Near-infrared photometry of planetary nebulae (PNe) allows the classification of those objects (Whitelock 1985, Pe\~na \& Torres-Peimbert 1987). We present the largest homogeneous sample.}

 The DENIS imaging survey (Epchtein et al.\ 1997) gives a nearly complete overview of the southern sky in the three NIR bands Gunn-I, J, and K$_{\rm s}$. 
The images are taken simultaneously in all three bands, which leads to a very high accuracy in the colors of the objects independent from  photometric errors. Photometry on the high resolution images allowed us a much better removal of the stellar background than the aperture photometries done in the past.
The 135 objects presented here form the largest homogeneous sample of NIR photometries of PNe so far.
For the calibration of our photometries we used the DENIS online zero points, 
taking into account a small offset to the values derived at the Paris Data Analysis Center. The calculated magnitudes were dereddened using the extinction constants from Tylenda et al.\ (1992).
Distances and linear radii were calculated from the 5~GHz flux using the method of Schneider \& Buckley (1996).
About one third of the objects overlap with measurements in the literature in the J and K band (no I band photometry exists up to now). The comparison shows clearly the expected effect:
While the brighter objects correspond very well, about 30\%
of the fainter ones are systematically brighter in the older aperture photometries (Fig.\ 1). This is caused by uncleaned stellar background in those works. Thus, the effect is significantly stronger in K than in J.
Especially the values from Persi et al.\ (1997) suffer from this effect.
We also find a correlation with the galactic longitude: The deviations increase towards the bulge.
As already pointed out in the literature, there is a clear correlation of the K band photometry with the radio fluxes.
In the J band, the correlation is not as good. Our J values are lower than expected, which is in clear contradiction to the results of Whitelock (1985). 
This effect is even stronger in the I Band.
These and other results will be discussed more detailed in a forthcoming paper.

\acknowledgements This project was supported by the FWF project P11675-AST, and by the BMBWK, Sektion VIII/A/5. Support from the organizers to participate in the conference is gratefully acknowledged.

\begin{figure}[!t]
\includegraphics[width=\columnwidth]{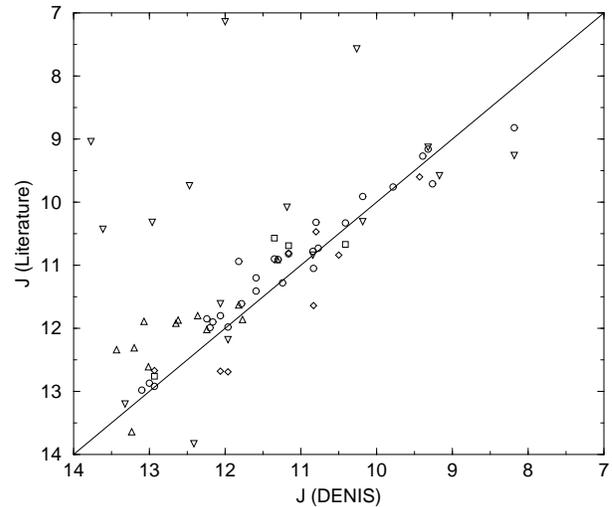}

  \caption{Comparison of our measured J magnitudes to values from the literature: Whitelock 1985 (circles), Pe\~na \& Torres-Peimbert 1987 (squares), Persi et al.\ 1987 (triangles down), Preite-Martinez \& Persi 1989 (triangles up), and Philips \& Cuesta 1994 (diamonds). 
}
    \label{fig:diag}
\end{figure}

\newpage

\begin{table*}[H]
\caption{Catalogue of NIR photometries of planetary nebulae.}
\begin{center}
{\tiny
\begin{tabular}{ l l r r r r r r r c r r }
\hline\hline
PN G & Name & \multicolumn{1}{c}{I} & \multicolumn{1}{c}{J} & \multicolumn{1}{c}{K$\rm _s$} & \multicolumn{1}{c}{$E_{\rm B-V}$} & (I-J)$_0$ & (J-K)$_0$ & 5 GHz  & Ref. & Distance & Radius \\
 &  &  &  &  &  &  &  & [mJy] &  &  \multicolumn{1}{c}{[kpc]} & \multicolumn{1}{c}{[pc]} \\
\hline
000.0$-$06.8 & H 1-62 & 13.62 & 13.08 & 12.96 &  &  &  &  &  &  & \\
000.1$+$04.3 & H 1-16 & 15.41 & 13.16 & 11.79 & 1.50 & 1.30 & 0.57 &  &  &  & \\
000.3$+$12.2 & IC 4634 & 12.25 & 11.35 & 10.76 & 0.25 & 0.74 & 0.45 & 100 & 2 & 3.14 & 0.064\\
000.3$-$04.6 & M 2-28 & 15.47 & 13.86 & 12.89 & 0.83 & 1.09 & 0.53 & 10 & 1 & 7.82 & 0.091\\
000.7$+$04.7 & H 2-11 & 16.45 & 13.94 & 12.38 & 2.15 & 1.15 & 0.41 & 28 & 4 & 7.09 & \\
001.7$+$05.7 & H 1-14 & 15.86 & 13.97 & 13.06 & 1.24 & 1.11 & 0.24 & 22 & 1 & 5.44 & 0.087\\
002.1$-$02.2 & M 3-20 & 15.42 & 13.65 & 14.05 & 0.88 & 1.22 & $-$0.87 & 40 & 1 & 4.56 & 0.073\\
002.2$-$02.7 & M 2-23 & 13.42 & 11.97 & 11.68 & 0.45 & 1.17 & 0.05 & 41 & 2 & 4.09 & 0.084\\
002.4$+$05.8 & NGC 6369 & 11.43 & 9.43 & 7.91 & 1.26 & 1.21 & 0.84 & 2002 & 1,2,3 & 0.70 & 0.064\\
002.6$+$08.1 & H 1-11 & 15.03 & 13.65 & 12.79 & 0.70 & 0.94 & 0.48 & 13 & 1,2 & 6.40 & 0.099\\
002.9$+$06.5 & PM 1-149 & 14.68 & 13.85 & 13.21 &  &  &  & 3.6 & 4 & 11.26 & \\
003.2$-$06.2 & M 2-36 & 13.95 & 12.96 & 12.62 & 0.30 & 0.80 & 0.18 & 25 & 1 & 4.81 & 0.094\\
003.4$-$04.8 & H 2-43 & 14.95 & 11.66 & 8.60 & 0.86 & 2.75 & 2.60 & 25 & 1 & 4.59 & 0.100\\
003.9$-$14.9 & Hb 7 & 13.11 & 12.42 & 11.99 & 0.19 & 0.57 & 0.33 & 30 & 1 & 6.03 & 0.058\\
004.0$-$03.0 & M 2-29 & 14.51 & 13.70 & 12.42 & 0.58 & 0.45 & 0.97 & 8 & 1,2 & 9.41 & 0.082\\
004.2$-$04.3 & H 1-60 & 14.47 & 13.56 & 12.93 & 0.42 & 0.65 & 0.41 &  &  &  & \\
004.6$+$06.0 & H 1-24 & 15.67 & 14.01 & 13.49 & 0.98 & 1.05 & 0.00 & 15 & 1,2 & 5.39 & 0.112\\
004.8$-$22.7 & He 2-436 & 15.72 & 14.31 & 12.71 & 0.39 & 1.16 & 1.39 & 23 & 1,3 & 4.48 & 0.109\\
005.1$-$03.0 & H 1-58 & 14.42 & 13.60 & 12.12 & 1.50 & $-$0.13 & 0.68 &  &  &  & \\
006.1$+$08.3 & M 1-20 & 14.07 & 12.65 & 11.24 & 0.70 & 0.98 & 1.02 & 51 & 1,2 & 4.14 & 0.070\\
007.2$+$01.8 & Hb 6 & 13.13 & 11.00 & 10.10 & 1.48 & 1.20 & 0.11 & 243 & 2 & 2.77 & 0.034\\
007.8$-$04.4 & H 1-65 & 13.51 & 12.86 & 12.25 & 0.64 & 0.25 & 0.27 & 17 & 1 & 5.38 & 0.104\\
008.3$-$01.1 & M 1-40 & 14.15 & 11.69 & 10.46 & 1.76 & 1.35 & 0.28 & 208 & 1,2 & 2.93 & 0.035\\
009.3$+$04.1 & Th 4-6 & 16.05 & 14.27 & 13.39 & 0.65 & 1.37 & 0.53 & 6 & 4 & 8.85 & \\
009.4$-$05.0 & NGC 6629 & 11.36 & 10.50 & 9.82 & 0.63 & 0.46 & 0.34 & 275 & 2 & 1.82 & 0.068\\
010.1$+$00.7 & NGC 6537 & 12.44 & 10.80 & 9.46 & 1.48 & 0.71 & 0.54 & 624 & 1 & 1.62 & 0.039\\
010.6$+$03.2 & Th 4-10 & 15.88 & 13.92 & 14.00 & 0.85 & 1.42 & $-$0.53 &  &  &  & \\
010.7$-$06.7 & Pe 1-13 & 14.94 & 13.96 & 13.36 & 0.46 & 0.69 & 0.36 & 3 & 1,2 & 8.60 & 0.158\\
010.8$-$01.8 & NGC 6578 & 12.82 & 9.39 & 7.38 & 0.98 & 2.81 & 1.49 & 170 & 2,3 & 2.64 & 0.054\\
011.0$-$05.1 & M 1-47 & 14.35 & 13.70 & 13.85 & 0.23 & 0.51 & $-$0.27 & 14 & 1,2 & 7.22 & 0.080\\
011.1$-$07.9 & SB 17 & 11.33 & 10.53 & 7.49 &  &  &  &  &  &  & \\
011.3$-$09.4 & H 2-48 & 11.94 & 11.54 & 10.98 & 0.49 & 0.09 & 0.29 & 66 & 1,2 & 5.71 & 0.028\\
011.7$-$00.0 & M 1-43 & 13.66 & 12.78 & 11.69 & 1.82 & $-$0.26 & 0.11 &  &  &  & \\
011.9$+$04.2 & M 1-32 & 13.22 & 11.74 & 10.39 & 1.04 & 0.83 & 0.79 & 61 & 2 & 3.80 & 0.070\\
012.2$+$04.9 & PM 1-188 & 15.02 & 13.15 & 8.78 &  &  &  &  &  &  & \\
013.1$+$04.1 & M 1-33 & 13.79 & 12.30 & 11.29 & 1.04 & 0.83 & 0.45 & 60 & 1,2 & 4.52 & 0.053\\
014.2$+$04.2 & Sa 3-111 & 14.59 & 10.36 & 8.28 & 1.82 & 3.07 & 1.10 &  &  &  & \\
014.9$+$06.4 & K 2-5 & 10.01 & 8.67 & 7.80 & 0.98 & 0.72 & 0.35 &  &  &  & \\
015.4$-$04.5 & M 1-53 & 12.52 & 11.34 & 10.39 & 0.72 & 0.72 & 0.56 & 53 & 1 & 4.34 & 0.063\\
015.9$+$03.3 & M 1-39 & 14.18 & 12.03 & 10.83 & 1.82 & 1.01 & 0.22 & 98 & 2 & 3.79 & 0.046\\
016.0$-$04.3 & M 1-54 & 13.81 & 12.91 & 11.73 & 0.64 & 0.50 & 0.83 & 38 & 1 & 3.47 & 0.109\\
018.9$+$03.6 & M 4-8 & 14.84 & 13.46 & 12.45 &  &  &  & 19 & 1,2 & 10.11 & \\
019.4$-$05.3 & M 1-61 & 12.46 & 11.43 & 10.50 & 0.73 & 0.57 & 0.54 & 97 & 1,2 & 5.08 & \\
019.7$+$03.2 & M 3-25 & 14.73 & 12.69 & 11.20 & 1.78 & 0.92 & 0.53 & 76 & 1,2 & 4.48 & 0.042\\
019.7$-$04.5 & M 1-60 & 13.94 & 12.36 & 11.51 & 1.07 & 0.91 & 0.27 & 48 & 2 & 3.65 & 0.088\\
019.8$+$05.6 & CTS 1 & 15.61 & 13.83 & 12.78 & 1.52 & 0.81 & 0.24 &  &  &  & \\
020.7$-$05.9 & Sa 1-8 & 13.60 & 13.09 & 12.75 & 0.57 & 0.15 & 0.04 & 11 & 1,2 & 6.05 & 0.117\\
022.0$-$03.1 & M 1-58 & 13.87 & 12.78 & 11.92 & 0.85 & 0.56 & 0.41 & 60 & 1 & 4.07 & 0.063\\
022.1$-$02.4 & M 1-57 & 13.79 & 12.51 & 11.43 & 1.26 & 0.48 & 0.40 & 70 & 2 & 3.50 & 0.071\\
022.5$+$01.0 & MaC 1-13 & 15.87 & 14.11 & 11.84 & 1.30 & 0.94 & 1.57 &  &  &  & \\
023.9$-$02.3 & M 1-59 & 13.32 & 11.83 & 10.89 & 1.11 & 0.79 & 0.35 & 108 & 2 & 3.77 & 0.042\\
025.9$-$02.1 & Pe 1-15 & 13.63 & 12.65 & 11.67 & 1.02 & 0.33 & 0.44 & 8 & 2 & 8.18 & 0.099\\
025.9$-$10.9 & Na 2 & 16.05 & 15.29 & 13.67 & 0.59 & 0.39 & 1.30 &  &  &  & \\
027.3$-$02.1 & Pe 1-18 & 15.04 & 12.76 & 12.29 & 1.95 & 1.05 & $-$0.58 & 42 & 1,2 & 4.44 & 0.073\\
027.7$+$00.7 & M 2-45 & 15.26 & 12.41 & 11.10 & 2.08 & 1.54 & 0.18 & 154 & 2 & 3.00 & 0.047\\
028.5$+$01.6 & M 2-44 & 14.04 & 12.47 & 10.48 & 1.27 & 0.78 & 1.30 & 54 & 1,2 & 3.98 & 0.071\\
028.5$+$05.1 & K 3-2 & 15.31 & 13.32 & 12.25 & 2.15 & 0.64 & $-$0.09 & 31 & 1,2 & 4.13 & 0.100\\
029.2$-$05.9 & NGC 6751 & 12.59 & 11.96 & 10.86 & 0.20 & 0.50 & 1.00 & 63 & 1,2,3 & 2.45 & 0.122\\
032.5$-$03.2 & K 3-20 & 14.48 & 13.12 & 11.33 & 1.30 & 0.55 & 1.09 &  &  &  & \\
032.9$-$02.8 & K 3-19 & 15.61 & 13.67 & 12.69 & 1.50 & 1.00 & 0.17 & 23 & 1,2 & 9.64 & \\
034.5$-$06.7 & NGC 6778 & 12.83 & 11.96 & 11.50 & 0.38 & 0.64 & 0.25 & 55 & 1,2,3 & 2.87 & 0.110\\
206.4$-$40.5 & NGC 1535 & 11.28 & 10.83 & 10.51 & 0.07 & 0.40 & 0.29 & 166 & 1 & 1.86 & 0.095\\
221.3$-$12.3 & IC 2165 & 12.21 & 11.16 & 10.35 & 0.47 & 0.75 & 0.56 & 202 & 1 & 2.44 & 0.053\\
226.4$-$03.7 & PB 1 & 14.55 & 13.61 & 13.32 & 0.85 & 0.40 & $-$0.17 & 18 & 1,3 & 4.78 & 0.116\\
232.8$-$04.7 & M 1-11 & 12.25 & 10.84 & 8.99 & 0.97 & 0.80 & 1.33 & 113 & 2 & 4.57 & \\
234.8$+$02.4 & NGC 2440 & 11.21 & 10.41 & 9.68 & 0.44 & 0.53 & 0.50 & 370 & 2 & 1.64 & 0.063\\
234.9$-$01.4 & M 1-14 & 13.02 & 12.00 & 11.35 & 0.57 & 0.66 & 0.35 & 60 & 2 & 4.55 & \\
235.3$-$03.9 & M 1-12 & 13.08 & 12.06 & 10.54 & 0.55 & 0.67 & 1.23 & 41 & 2 & 6.99 & \\
242.6$-$11.6 & M 3-1 & 13.82 & 13.07 & 13.47 & 0.24 & 0.60 & $-$0.53 & 24 & 1,2 & 4.20 & 0.114\\
248.8$-$08.5 & M 4-2 & 14.57 & 13.72 & 12.69 & 0.24 & 0.70 & 0.90 & 19 & 1,2 & 5.83 & 0.088\\
249.0$+$06.9 & SaSt 1-1 & 11.85 & 11.62 & 11.42 &  &  &  &  &  &  & \\
253.9$+$05.7 & M 3-6 & 12.26 & 11.59 & 11.21 & 0.42 & 0.41 & 0.16 & 75 & 2 & 3.46 & 0.069\\
258.1$-$00.3 & He 2-9 & 13.43 & 11.59 & 10.12 & 1.41 & 0.95 & 0.71 & 170 & 3 & 3.26 & \\
272.1$+$12.3 & NGC 3132 & 9.58 & 9.26 & 8.53 & 0.10 & 0.26 & 0.68 & 230 & 1 & 1.45 & 0.105\\
274.1$+$02.5 & He 2-34 & 13.03 & 9.17 & 4.28 & 1.82 & 2.72 & 3.91 &  &  &  & \\
274.6$+$02.1 & He 2-35 & 14.63 & 13.56 & 13.35 & 0.59 & 0.70 & $-$0.11 & 24 & 1 & 5.94 & 0.072\\
279.6$-$03.1 & He 2-36 & 10.52 & 9.78 & 9.41 &  &  &  & 90 & 2 & 2.16 & 0.115\\
261.0$+$32.0 & NGC 3242 & 10.02 & 9.02 & 8.59 & 0.20 & 0.87 & 0.32 & 835 & 1 & 1.07 & 0.065\\
285.6$-$02.7 & He 2-47 & 12.07 & 10.77 & 10.11 & 0.53 & 0.96 & 0.38 & 170 & 1 & 3.14 & 0.038\\
286.3$-$04.8 & NGC 3211 & 12.91 & 12.20 & 11.78 & 0.22 & 0.57 & 0.30 & 80 & 1 & 2.57 & 0.100\\
\hline\hline
\end{tabular}
}
\end{center}
\end{table*}

\newpage

\begin{table*}[H]
\addtocounter{table}{-1}
\caption{(continued)}
\begin{center}
{\tiny
\begin{tabular}{ l l r r r r r r r c r r }
\hline\hline
PN G & Name & \multicolumn{1}{c}{I} & \multicolumn{1}{c}{J} & \multicolumn{1}{c}{K$\rm _s$} & \multicolumn{1}{c}{$E_{\rm B-V}$} & (I-J)$_0$ & (J-K)$_0$ & 5 GHz  & Ref. & Distance & Radius \\
 &  &  &  &  &  &  &  & [mJy] &  &  \multicolumn{1}{c}{[kpc]} & \multicolumn{1}{c}{[pc]} 
  \\  
\hline

292.8$+$01.1 & He 2-67 & 14.39 & 13.00 & 12.39 & 0.71 & 0.93 & 0.23 & 41 & 1,3 & 5.03 & 0.061\\
294.6$+$04.7 & NGC 3918 & 13.09 & 9.32 & 8.78 & 0.26 & 3.61 & 0.40 & 859 & 2,3 & 1.18 & 0.054\\
296.3$-$03.0 & He 2-73 & 13.87 & 12.24 & 11.44 & 0.92 & 1.05 & 0.31 & 76 & 1,2,3 & 4.44 & 0.043\\
307.2$-$09.0 & He 2-97 & 13.47 & 12.16 & 11.37 & 0.39 & 1.06 & 0.57 & 30 & 1,3 & 5.55 & 0.067\\
307.5$-$04.9 & MyCn 18 & 12.39 & 11.24 & 10.64 & 1.19 & 0.40 & $-$0.04 & 106 & 1,2 & 3.96 & 0.038\\
311.4$+$02.8 & He 2-102 & 14.15 & 13.10 & 12.48 & 0.82 & 0.53 & 0.17 & 33 & 1,2 & 4.25 & 0.093\\
320.3$-$28.8 & He 2-434 & 13.47 & 13.16 & 12.59 & 0.17 & 0.20 & 0.48 &  &  &  & \\
320.9$+$02.0 & He 2-117 & 13.52 & 11.31 & 9.90 & 1.76 & 1.10 & 0.46 & 267 & 2 & 2.68 & 0.032\\
321.3$+$02.8 & He 2-115 & 13.55 & 11.78 & 10.45 & 1.56 & 0.78 & 0.49 & 156 & 1,2 & 3.74 & 0.027\\
322.4$-$00.1 & Pe 2-8 & 14.93 & 11.82 & 10.05 & 2.60 & 1.47 & 0.37 & 100 & 1,2 & 5.16 & 0.020\\
324.2$+$02.5 & He 2-125 & 14.58 & 13.46 & 12.49 & 1.19 & 0.36 & 0.34 &  &  &  & \\
325.0$+$03.2 & He 2-129 & 14.79 & 13.07 & 11.41 & 1.30 & 0.90 & 0.96 & 35 & 1,2 & 7.65 & 0.030\\
325.8$+$04.5 & He 2-128 & 14.00 & 12.96 & 12.06 & 0.77 & 0.55 & 0.49 & 40 & 1,2 & 5.07 & 0.061\\
327.1$-$02.2 & He 2-142 & 12.18 & 10.26 & 8.80 & 1.13 & 1.21 & 0.86 & 65 & 1,2,3 & 4.84 & 0.042\\
327.8$+$10.0 & NGC 5882 & 11.11 & 10.18 & 9.54 & 0.25 & 0.76 & 0.51 & 334 & 1,2 & 1.78 & 0.060\\
327.8$-$01.6 & He 2-143 & 14.96 & 12.36 & 11.08 & 2.08 & 1.29 & 0.15 & 120 & 1,2 & 3.49 & 0.044\\
327.8$-$06.1 & He 2-158 & 14.19 & 13.77 & 12.94 & 0.33 & 0.21 & 0.65 &  &  &  & \\
330.7$+$04.1 & Cn 1-1 & 10.02 & 8.93 & 7.56 & 0.66 & 0.67 & 1.02 &  &  &  & \\
331.0$-$02.7 & He 2-157 & 14.26 & 13.20 & 12.10 & 1.04 & 0.40 & 0.54 & 30 & 1 & 6.67 & 0.048\\
331.7$-$01.0 & Mz 3 & 11.32 & 8.18 & 4.33 & 1.37 & 2.28 & 3.11 & 649 & 1,2,3 & 1.16 & 0.070\\
332.2$+$03.5 & Wray 16-199 & 15.20 & 13.26 & 12.16 &  &  &  &  &  &  & \\
332.9$-$09.9 & He 3-1333 & 10.17 & 9.49 & 6.67 & 0.59 & 0.31 & 2.50 &  &  &  & \\
334.3$-$09.3 & IC 4642 & 13.23 & 12.62 & 12.46 & 0.33 & 0.40 & $-$0.02 & 60 & 1,2 & 2.75 & 0.110\\
334.8$-$07.4 & SaSt 2-12 & 10.46 & 9.71 & 9.29 & 0.39 & 0.50 & 0.21 &  &  &  & \\
336.2$+$01.9 & Pe 1-6 & 14.95 & 13.23 & 12.29 & 1.37 & 0.86 & 0.20 & 40 & 1,2 & 3.54 & 0.103\\
336.3$-$05.6 & He 2-186 & 14.24 & 13.01 & 11.87 & 0.61 & 0.85 & 0.82 & 21 & 1,2 & 7.49 & 0.054\\
336.9$+$08.3 & StWr 4-10 & 14.82 & 14.28 & 14.13 & 0.92 & $-$0.04 & $-$0.35 &  &  &  & \\
338.8$+$05.6 & He 2-155 & 13.56 & 12.47 & 11.77 & 0.64 & 0.68 & 0.36 & 70 & 1,2 & 2.79 & 0.098\\
340.9$-$04.6 & Sa 1-5 & 16.39 & 15.06 & 14.12 & 1.04 & 0.67 & 0.38 &  &  &  & \\
342.1$+$27.5 & Me 2-1 & 13.72 & 12.93 & 12.37 & 0.08 & 0.74 & 0.52 & 30 & 2 & 5.17 & 0.075\\
342.8$-$06.6 & Cn 1-4 & 13.52 & 13.15 & 12.75 & 0.29 & 0.19 & 0.24 &  &  &  & \\
343.4$+$11.9 & H 1-1 & 15.74 & 14.87 & 13.69 & 0.35 & 0.66 & 0.99 &  &  &  & \\
343.5$-$07.8 & PC 17 & 13.83 & 13.21 & 12.92 & 0.47 & 0.32 & 0.04 &  &  &  & \\
345.0$+$03.4 & Vd 1-4 & 16.24 & 14.44 & 12.67 & 0.72 & 1.34 & 1.39 &  &  &  & \\
345.9$-$11.2 & ESO 279-14 & 13.30 & 12.31 & 11.43 &  &  &  &  &  &  & \\
347.7$+$02.0 & Vd 1-8 & 15.65 & 13.27 & 11.74 & 1.82 & 1.24 & 0.55 &  &  &  & \\
348.8$-$09.0 & He 2-306 & 14.24 & 13.66 & 12.93 & 0.29 & 0.41 & 0.57 &  &  &  & \\
350.9$+$04.4 & H 2-1 & 11.94 & 11.18 & 10.20 & 0.68 & 0.33 & 0.61 & 61 & 2 & 4.26 & 0.058\\
351.1$+$04.8 & M 1-19 & 14.27 & 13.00 & 12.37 & 0.79 & 0.77 & 0.20 & 26 & 1,2 & 4.78 & 0.093\\
352.8$-$00.2 & H 1-13 & 13.72 & 10.87 & 9.24 & 2.21 & 1.45 & 0.44 & >620 & 2 &  & \\
352.9$-$07.5 & Fg 3 & 12.08 & 11.30 & 10.60 & 0.46 & 0.49 & 0.46 & 107 & 1 & 4.77 & 0.023\\
353.3$+$06.3 & M 2-6 & 14.47 & 13.48 & 12.76 & 0.53 & 0.66 & 0.43 & 17 & 4 & 5.38 & \\
354.2$+$04.3 & M 2-10 & 14.78 & 13.25 & 12.62 & 1.09 & 0.84 & 0.05 & 9.1 & 2 & 8.68 & 0.084\\
355.2$-$02.5 & H 1-29 & 14.76 & 13.47 & 12.54 & 1.00 & 0.67 & 0.39 &  &  &  & \\
355.9$-$04.2 & M 1-30 & 13.74 & 12.82 & 12.02 & 0.64 & 0.52 & 0.45 & 31 & 2 & 5.49 & 0.067\\
356.2$-$04.4 & Cn 2-1 & 13.33 & 11.77 & 10.87 & 0.47 & 1.27 & 0.65 & 49 & 2 & 6.05 & 0.035\\
356.5$-$03.9 & H 1-39 & 14.30 & 13.34 & 12.63 & 1.02 & 0.32 & 0.17 & 13 & 2 & 7.12 & 0.086\\
357.1$+$01.2 & K 6-2 & 15.17 & 12.10 & 10.47 &  &  &  &  &  &  & \\
357.2$+$07.4 & M 4-3 & 15.10 & 13.43 & 12.76 & 0.98 & 1.06 & 0.15 & 28 & 2 & 4.68 & 0.091\\
357.4$-$03.5 & M 2-18 & 14.45 & 12.77 & 11.62 & 0.87 & 1.13 & 0.69 & 17 & 1,2 & 6.59 & 0.080\\
358.2$+$03.6 & M 3-10 & 14.57 & 12.95 & 11.93 & 1.07 & 0.95 & 0.44 & 29 & 1 & 6.60 & 0.051\\
358.3$+$01.2 & Bl B & 15.79 & 13.29 & 11.43 & 3.26 & 0.44 & 0.11 &  &  &  & \\
358.3$-$02.5 & Al 2-O & 14.40 & 12.46 & 10.66 &  &  &  & 37 & 4 & 6.68 & \\
358.3$-$21.6 & IC 1297 & 12.86 & 12.03 & 10.98 & 0.12 & 0.74 & 0.99 &  &  &  & \\
358.8$+$00.0 & Te 2022 & 12.06 & 10.10 & 8.61 &  &  &  &  &  &  & \\

\hline\hline
\end{tabular}
}
\\
\vspace{1mm}

\parbox{13cm}{
\tiny \noindent References:\ (1) Cahn, J.H., Kaler, J.B., \& Stanghellini, L. 1992, A\&AS, 94, 399\\
\phantom{References:}\ (2) Van de Steene,
G.C., \& Zijlstra, A.A. 1994, A\&AS, 108, 485 \\
\phantom{References:}\ (3) Acker, A., Ochsenbein, F., Stenholm, B., et.al. 1992, Strasbourg-ESO Catalogue of Galactic PNe \\
\phantom{References:}\ (4) Milne, D.K., \& Aller, L.H. 1975, A\&A, 38, 183
}
\vspace{5cm}
\phantom{X}
\end{center}
\end{table*}

\end{document}